\documentclass[12pt]{article}
\usepackage{amsmath}
\usepackage{graphicx}
\usepackage{bm}
\usepackage{fancyhdr}
\usepackage{amssymb}
\usepackage{caption}
\usepackage{subfigure}
\usepackage{setspace}
\begin{document}



\def\a{\alpha}
\def\b{\beta}
\def\d{\delta}
\def\e{\epsilon}
\def\g{\gamma}
\def\h{\mathfrak{h}}
\def\k{\kappa}
\def\l{\lambda}
\def\o{\omega}
\def\p{\wp}
\def\r{\rho}
\def\t{\tau}
\def\s{\sigma}
\def\z{\zeta}
\def\x{\xi}
\def\V={{{\bf\rm{V}}}}
 \def\A{{\cal{A}}}
 \def\B{{\cal{B}}}
 \def\C{{\cal{C}}}
 \def\D{{\cal{D}}}
\def\K{{\cal{K}}}
\def\O{\Omega}
\def\R{\bar{R}}
\def\T{{\cal{T}}}
\def\L{\Lambda}
\def\f{E_{\tau,\eta}(sl_2)}
\def\E{E_{\tau,\eta}(sl_n)}
\def\Zb{\mathbb{Z}}
\def\Cb{\mathbb{C}}

\def\R{\overline{R}}

\def\beq{\begin{equation}}
\def\eeq{\end{equation}}
\def\bea{\begin{eqnarray}}
\def\eea{\end{eqnarray}}
\def\ba{\begin{array}}
\def\ea{\end{array}}
\def\no{\nonumber}
\def\le{\langle}
\def\re{\rangle}
\def\lt{\left}
\def\rt{\right}

\newtheorem{Theorem}{Theorem}
\newtheorem{Definition}{Definition}
\newtheorem{Proposition}{Proposition}
\newtheorem{Lemma}{Lemma}
\newtheorem{Corollary}{Corollary}
\newcommand{\proof}[1]{{\bf Proof. }
        #1\begin{flushright}$\Box$\end{flushright}}

\baselineskip=20pt

\newfont{\elevenmib}{cmmib10 scaled\magstep1}
\newcommand{\preprint}{
   \begin{flushleft}
   \end{flushleft}\vspace{-1.3cm}
   \begin{flushright}\normalsize
   \end{flushright}}
\newcommand{\Title}[1]{{\baselineskip=26pt
   \begin{center} \Large \bf #1 \\ \ \\ \end{center}}}
   \newcommand{\Author}{\begin{center}
   \large \bf
Pei Sun${}^{a,b}$, Zhi-Rong Xin${}^{a,b}$, Yi Qiao${}^{a,b,c}$,  Kun Hao${}^{a,b}$, Like Cao${}^{b,d}$,
Junpeng Cao${}^{c,e,f}\footnote{Corresponding author: junpengcao@iphy.ac.cn}$, Tao Yang${}^{a,b,d}$ and Wen-Li Yang${}^{a,b,d}\footnote{Corresponding author: wlyang@nwu.edu.cn}$
 \end{center}}
\newcommand{\Address}{\begin{center}
     ${}^a$ Institute of Modern Physics, Northwest University,
     Xian 710127, China\\
     ${}^b$ Shaanxi Key Laboratory for Theoretical Physics Frontiers,  Xian 710127, China\\
     ${}^c$ Beijing National Laboratory for Condensed Matter
           Physics, Institute of Physics, Chinese Academy of Sciences, Beijing
           100190, China\\
     ${}^d$ School of Physics, Northwest University,  Xian 710127, China\\
     ${}^e$Songshan Lake Materials Laboratory, Dongguan, Guangdong 523808, China \\
     ${}^f$School of Physical Sciences, University of Chinese Academy of
Sciences, Beijing, China\\
   \end{center}}

\preprint \thispagestyle{empty}
\bigskip\bigskip\bigskip

\Title{Surface energy and elementary excitations of the XXZ
spin chain with arbitrary boundary fields} \Author

\Address \vspace{0.1cm}

\vspace{1truecm}

\begin{abstract}
The thermodynamic properties of the XXZ spin chain with integrable open boundary conditions at the gaped region (i.e., the anisotropic parameter $\eta$ being a real number)
are investigated.
It is shown that the contribution of the inhomogeneous term in the $T-Q$ relation of the ground state and elementary excited state can be neglected when the size of the system $N$ tends to infinity. The surface energy and elementary excitations induced by the unparallel boundary magnetic fields are obtained.
\vspace{1truecm}


\end{abstract}

\newpage

\hbadness=10000

\tolerance=10000

\hfuzz=150pt

\vfuzz=150pt
\section{Introduction}
\setcounter{equation}{0}

Since  Yang and Baxter's pioneering works~\cite{1cn Yang, 3Baxter 2}, the exactly solvable quantum
systems have attracted a great deal of interest because they can provide us solid benchmarks for
understanding the many-body effects. Especially the exact solutions are very important in nano-scale systems
where alternative approaches involving mean field approximations or perturbations have failed. 
At present, the integrable models have many applications in
statistical physics, low-dimensional condensed matter physics \cite{Tak99}, and even some mathematical areas such as quantum groups and quantum algebras.

The coordinate Bethe ansatz \cite{12F.C. Alcaraz} and the algebraic Bethe ansatz \cite{Kor93,13E.K. Sklyanin} are the standard methods to obtain the exact solutions of models with $U(1)$ symmetry.
However, when the $U(1)$ symmetry is broken, these methods cannot be directly applied to due to lacking the reference states.
Then the off-diagonal Bethe ansatz (ODBA) was proposed to study the models with or without $U(1)$ symmetry \cite{14J.Cao1,15J.Cao2,16J.Cao3,17J.Cao4,19J.Cao6}.
For further information, we refer the reader to the reference \cite{22Y.Wang}.

In this paper, we consider the open spin-$\frac{1}{2}$ XXZ quantum spin chain with nondiagonal boundary terms, which is given by the Hamiltonian
\begin{eqnarray}
  H=\sum_{j=1}^{N-1}[\sigma_j^x\sigma_{j+1}^x+\sigma_j^y\sigma_{j+1}^y+
  \cosh\eta\,\sigma_j^z\sigma_{j+1}^z]+\vec{h}_1.\vec{\sigma}_1+
  \vec{h}_N.\vec{\sigma}_N,
 \label{Hamiltion0}
\end{eqnarray}
where $\sigma_j^\alpha (\alpha=x,y,z)$ are the Pauli matrices at site $j$ and
$\{\vec{h}_i=(h_i^x,h_i^y,h_i^z)|i=1,N\}$ are two boundary magnetic fields, $\eta$ is the so-called anisotropic parameter. This is a prototypical
integrable quantum spin chain with boundary fields. It can be related to many other models such as
the sine-Gordon field theory~\cite{boundary3}. Moreover, this model has applications in various branches
of physics, including condensed matter and statistical mechanics.

The Bethe ansatz solution of model (\ref{Hamiltion0}) with diagonal boundary fields has been known~\cite{12F.C. Alcaraz,13E.K. Sklyanin}.
If the boundary reflection have the off-diagonal elements, the eigenvalue and the eigenstates has been obtained by the off-diagonal Bethe ansatz~\cite{15J.Cao2,17J.Cao4,eigenstate,wenfakai1,wenfakai2}.
The eigenvalues of the system for arbitrary boundary fields is given by an inhomogeneous $T-Q$ relation, giving rise to the fact that the study of the
thermodynamic limit becomes more involved in.  However, if the crossing parameter $\eta$ is an arbitrary imaginary number, there exist a series of infinite special points at which the inhomogeneous $T-Q$ relation reduces to the homogeneous one and thus the associated Bethe ansatz equations become the standard ones \cite{19J.Cao6}.
In the thermodynamic limit, these points become dense on the imaginary line which allows ones to study
the thermodynamics properties such as the ground state and the surface energy when the anisotropic parameter $\eta$ is an imaginary number (namely, the open XXZ chain at the gapless region) \cite{19J.Cao6}. However, if $\eta$ is an arbitrary real number, there does not exist the series points   and thus the previous analysis fails.

In this paper, we study the thermodynamic limit of the model (\ref{Hamiltion0}) with $\eta$ being an arbitrary real number under the nondiagonal
boundary fields. We first address the contribution of the inhomogeneous term with finite system-size.
It is shown that the contribution of the inhomogeneous term in the associated $T-Q$ relation to the ground state energy and elementary excitation can be neglected when
the system-size $N$ tends to infinity. Then based on the reduced Bethe ansatz equation, we study the surface energy~\cite{surface1,surface2,surface3} which contains the effects induced
by the unparallel boundary fields. Furthermore, we obtain the elementary excitation energy.

The paper is organized as follows. In section~\ref{sec:The model}, the exact solution of the model is briefly reviewed. In
section~\ref{sec:surface excitations},  we give the reduced homogeneous $T-Q$ relation and calculate
the surface excitations which comes from the boundary strings.
In section~\ref{sec:Finite size correction},
we focus on the contribution of the inhomogeneous term to the ground state
energy. In section~\ref{sec:surface energy}, we study the thermodynamic limit and surface energy of the model with $\eta$ being an arbitrary real number. In section~\ref{sec:Elementary excitation-anti}, we further calculate the elementary excitation energy induced by the boundary fields.
Section~\ref{sec:concluding remarks} gives some discussions.

\section{The model and its ODBA solution}
\label{sec:The model}

\setcounter{equation}{0}

In order to address the boundary reflection clearly, we rewrite the Hamiltonian \eqref{Hamiltion0} as
\begin{eqnarray}
\label{Hamiltonian}
  H &=& \sum_{j=1}^{N-1}[\sigma_{j}^{x}\sigma_{j+1}^{x}+\sigma_{j}^{y}\sigma_{j+1}^{y}+\cosh\eta\sigma_{j}^{z}\sigma_{j+1}^{z}]\nonumber\\
  &&+\frac{\sinh\eta}{\sinh\alpha_-\cosh\beta_-}(\cosh\alpha_-\sinh\beta_-\sigma_1^z+
  \cosh\theta_-\sigma_1^x+i\sinh\theta_-\sigma_1^y)\nonumber\\
 &&-\frac{\sinh\eta}{\sinh\alpha_+\cosh\beta_+}(\cosh\alpha_+\sinh\beta_+\sigma_N^z-
  \cosh\theta_+\sigma_N^x-i\sinh\theta_+\sigma_N^y),
\end{eqnarray}
where $\alpha_{\mp}$, $\beta_{\mp}$ and $\theta_{\mp}$ are the boundary parameters which parameterize the components of boundary fields and are related to
the parameters of the $K$-matrices (see (\ref{Kmatrixfu} and (\ref{K-1}) below). The integrability of the model is associated with the $R$-matrix
\begin{eqnarray}
  R_{0,j}(u) = \frac{1}{2}\left[\frac{\sinh(u+\eta)}{\sinh(\eta)}(1+\sigma_j^z
  \sigma_0^z)+\frac{\sinh u}{\sinh\eta}(1-\sigma_j^z\sigma_0^z)\right]
  +\frac{1}{2}(\sigma_j^x\sigma_0^x+\sigma_j^y\sigma_0^y),
\end{eqnarray}
where $u$ is the spectral parameter and $\eta$ is the bulk anisotropic parameter.
The $R$-matrix satisfies the Yang-Baxter equation (YBE)
\begin{eqnarray}
R_{12}(u_1-u_2)R_{13}(u_1-u_3)R_{23}(u_2-u_3)
=R_{23}(u_2-u_3)R_{13}(u_1-u_3)R_{12}(u_1-u_2).
\label{QYBE}
\end{eqnarray}
The boundary magnetic fields are described by the reflection matrix \cite{boundary3,boundary1}
\begin{eqnarray}
 K^-(u) &=& \left(
                \begin{array}{cc}
                  K_{11}^-(u) & K_{12}^-(u) \\
                  K_{21}^-(u) & K_{22}^-(u) \\
                \end{array}
              \right),\nonumber\\
 K_{11}^-(u) &=&
 2[\sinh(\alpha_-)\cosh(\beta_-)\cosh(u)+\cosh(\alpha_-)\sinh(\beta_-)\sinh(u)],\nonumber\\  K_{22}^-(u) &=&
 2[\sinh(\alpha_-)\cosh(\beta_-)\cosh(u)-\cosh(\alpha_-)\sinh(\beta_-)\sinh(u)],\nonumber\\
 K_{12}^-(u) &=& e^{\theta_-}\sinh(2u),\quad
 K_{21}^-(u) = e^{-\theta_-}\sinh(2u),
\label{Kmatrixfu}
\end{eqnarray}
and the dual reflection matrix
\begin{eqnarray}
K^+(u)=K^-(-u-\eta)\left|_{(\alpha_-,\beta_-,\theta_-)\rightarrow
(-\alpha_+,-\beta_+,\theta_+)}.\right.\label{K-1}
\end{eqnarray}
The former satisfies the reflection equation (RE)
\begin{eqnarray}
\label{kfu RKRK}
 &&R_{12}(u_1-u_2)K^-_1(u_1)R_{21}(u_1+u_2)K^-_2(u_2)\nonumber\\
 &&=
K^-_2(u_2)R_{12}(u_1+u_2)K^-_1(u_1)R_{21}(u_1-u_2),\label{QREK1}
\end{eqnarray}
and the latter satisfies the dual RE
\begin{eqnarray}
&&R_{12}(u_2-u_1)K^+_1(u_1)R_{21}(-u_1-u_2-2\eta)K^+_2(u_2)\no\\
&&= K^+_2(u_2)R_{12}(-u_1-u_2-2\eta)K^+_1(u_1)R_{21}(u_2-u_1).
\label{QREK2}
\end{eqnarray}
In order to show the intergrability of the system, we first
introduce the ``row-to-row" monodromy matrices $T_0(u)$
and $\hat{T}_0(u)$
\begin{eqnarray}
T_0(u)&=&R_{0N}(u-\theta_N)R_{0\,N-1}(u-\theta_{N-1})\cdots
R_{01}(u-\theta_1),\label{}\\
\hat{T}_0(u)&=&R_{10}(u+\theta_1)R_{20}(u+\theta_{2})\cdots
R_{N0}(u+\theta_N),
\end{eqnarray}
where $\{\theta_j, j=1, \cdots, N\}$ are the inhomogeneous
parameters. The one-row monodromy matrices are the $2\times 2$
matrices in the auxillary space $0$ and their elements act on the
quantum space ${\rm\bf V}^{\otimes N}$.
The transfer matrix of the system reads
\begin{eqnarray}
t(u)=tr_0\{K^+_0(u)T_0(u)K^-_0(u)\hat{T}_0(u)\}.
\label{transfer matrix}
\end{eqnarray}
Using the YBE (\ref{QYBE}), RE (\ref{QREK1}) and dual RE (\ref{QREK2}),
one can prove that the transfer matrices with different spectral parameters
commute with each other, namely, $[t(u), t(v)]=0$. Therefore, $t(u)$ serves
as the generating function of all the conserved quantities of the
system. The model Hamiltonian~\eqref{Hamiltonian} is constructed by taking the
derivative of the logarithm of the transfer matrix
\begin{eqnarray}
&&H=\sinh\eta \frac{\partial \ln t(u)}{\partial
u}|_{u=0,\{\theta_j\}=0}-N\cosh(\eta)-\tanh\eta\sinh\eta. \label{hamilton}
\end{eqnarray}
By using the off-diagonal Bethe ansatz method, the eigenvalue $\Lambda(u)$ of the
transfer matrix $t(u)$ can be given by the inhomogenous $T-Q$ relation~\cite{15J.Cao2},
\begin{eqnarray}
\label{T-Q}
  \Lambda(u) &=& a(u)\frac{Q(u-\eta)}{Q(u)}+d(u)\frac{Q(u+\eta)}{Q(u)}\nonumber \\
   &&+
   \frac{2c\sinh(2u)\sinh(2u+2\eta)}{Q(u)}\bar{A}(u)\bar{A}(-u-\eta),
\end{eqnarray}
where
\begin{eqnarray}
  &&c = \cosh\left[(N+1)\eta+\alpha_-+\beta_-+\alpha_++\beta_+\right]
  -\cosh(\theta_--\theta_+), \no  \\
  &&\bar{A}(u)=\prod_{l=1}^N\frac{\sinh(u-\theta_l+\eta)\sinh(u+\theta_l+\eta)}{\sinh^2\eta}, \no \\
  &&Q(u) = \prod_{j=1}^{N}\frac{\sinh(u-u_j)\sinh(u+u_j+\eta)}{\sinh^2\eta},
 \no \\
&&  a(u) =d(-u-\eta)= -4\frac{\sinh(2u+2\eta)}{\sinh(2u+\eta)}\sinh(u-\alpha_-)\sinh(u-\alpha_+)\nonumber\\
  && \qquad \qquad\times\cosh(u-\beta_-)\cosh(u-\beta_+)
  \bar{A}(u).
\end{eqnarray}
The $N$ Bethe roots $\{u_j|j=1,\dots,N\}$ should satisfy the Bethe ansatz equations (BAEs)
\begin{eqnarray}
 \label{BAEs}
 &&a(u_j)Q(u_j-\eta)+d(u_j)Q(u_j+\eta)\nonumber\\
 &&\qquad\qquad +2c\sinh(2u_j)\sinh(2u_j+2\eta)\bar{A}(u_j)\bar{A}(-u_j-\eta)=0.
 \end{eqnarray}
The eigenvalue of the Hamiltonian~\eqref{Hamiltonian} in terms of the Bethe roots is
\begin{eqnarray}
\label{eigenvalue1}
  E &=& -\sinh\eta[\coth\alpha_-+\tanh\beta_-+\coth\alpha_++\tanh\beta_+]\nonumber\\
  &&+2\sum_{j=1}^{N}\frac{\sinh^2\eta}{\sinh u_j\sinh(u_j+\eta)}+(N-1)\cosh\eta.
\end{eqnarray}

\section{Reduced $T-Q$ relation and surface excitations}
\label{sec:surface excitations}
\setcounter{equation}{0}

In order to study the contribution of the inhomogeneous term in (\ref{T-Q}), we first consider the following reduced $T-Q$ relation
\begin{eqnarray}
  \Lambda_{hom}(u)=a(u)\frac{Q(u-\eta)}{Q(u)}+d(u)\frac{Q(u+\eta)}{Q(u)}.\label{xexa}
\end{eqnarray}
We note that the non-diagonal boundary parameters are included in the above reduced $T-Q$ relation.
For convenience, we put $u_j=i\frac{\lambda_j}{2}-\frac{\eta}{2}$ with $\eta>0,\lambda_j\in(-\pi,\pi]$. From the singularity analysis of $\Lambda_{hom}(u)$, we obtain the reduced BAEs
\begin{eqnarray}
&&\left[\frac{\sin(\frac{\lambda_j}{2}-i\frac{\eta}{2})} {\sin(\frac{\lambda_j}{2}+i\frac{\eta}{2})}\right]^{2N} \frac{\sin(\lambda_j-i\eta)}{\sin(\lambda_j+i\eta)}
  \frac{\sin(\frac{\lambda_j}{2}+i\frac{\eta}{2}+i\alpha_+)}{\sin(\frac{\lambda_j}{2}-i\frac{\eta}{2}-i\alpha_+)}\nonumber\\[4pt]
&&\times\frac{\sin(\frac{\lambda_j}{2}+i\frac{\eta}{2}+i\alpha_-)}{\sin(\frac{\lambda_j}{2}-i\frac{\eta}{2}-i\alpha_-)}
  \frac{\cos(\frac{\lambda_j}{2}+i\frac{\eta}{2}+i\beta_+)}{\cos(\frac{\lambda_j}{2}-i\frac{\eta}{2}-i\beta_+)}
  \frac{\cos(\frac{\lambda_j}{2}+i\frac{\eta}{2}+i\beta_-)}{\cos(\frac{\lambda_j}{2}-i\frac{\eta}{2}-i\beta_-)}
  \nonumber\\[4pt]
&&\quad\quad =\prod_{l=1}^{M} \frac{\sin(\frac{\lambda_j-\lambda_l}{2}-i\eta)}{\sin(\frac{\lambda_j-\lambda_l}{2}+i\eta)}
\frac{\sin(\frac{\lambda_j+\lambda_l}{2}-i\eta)}{\sin(\frac{\lambda_j+\lambda_l}{2}+i\eta)}
,\qquad j=1, \dots, M. \label{BAEsre}
\end{eqnarray}
We define the reduced eigenvalues as
\begin{eqnarray}
\label{Ehom}
 E_{hom}&=&
 \sinh\eta \frac{\partial \ln \Lambda_{hom}(u)}{\partial
u}|_{u=0,\{\theta_j\}=0}-N\cosh(\eta)-\tanh\eta\,\sinh\eta \no \\
 &=& \sum_{j=1}^{M}\frac{4\sinh^2\eta}{\cos\lambda_j-\cosh\eta}+N \cosh\eta+E_0,
\end{eqnarray}
where
\begin{eqnarray}
E_0 =-\sinh\eta(\coth\alpha_-+\coth\alpha_++\tanh\beta_++\tanh\beta_-)-\cosh(\eta).
\end{eqnarray}
Taking the logarithm of BAEs~\eqref{BAEsre}, we obtain
\begin{eqnarray}
&& 2N\phi_1(\lambda_j)+\phi_2(2\lambda_j)-\phi_{(2\alpha_-/\eta+1)}(\lambda_j)-\phi_{(2\alpha_+/\eta+1)}(\lambda_j)
+\gamma_{+}(\lambda_j)+\gamma_{-}(\lambda_j)+\pi \nonumber\\
&&=2\pi I_j+\sum_{l=1}^{M}[\phi_2(\lambda_j-\lambda_l)+\phi_2(\lambda_j+\lambda_l)],\qquad j=1,\dots, M,\label{BAEsln}
\end{eqnarray}
with $I_j$ being an integer which determine the eigenvalue and
\begin{eqnarray}
  \phi_m(\lambda_j)=-i\ln\frac{\sin(\frac{\lambda_j}{2}-i\frac{m\eta}{2})}
  {\sin(\frac{\lambda_j}{2}+i\frac{m\eta}{2})}, \qquad
  \gamma_{\pm}(\lambda_j)=-i\ln\frac{\cos(\frac{\lambda_j}{2}+i\frac{\eta}{2}+i\beta_{\pm})}
  {\cos(\frac{\lambda_j}{2}-i\frac{\eta}{2}-i\beta_{\pm})}.
\end{eqnarray}
Define the counting function as $Z(\lambda_j)=\frac{I_j}{2N}$, then the BAEs~\eqref{BAEsln} read
\begin{eqnarray}
Z(\lambda) &=& \frac{1}{2\pi}
\Bigg\{\phi_1(\lambda)+\frac{1}{2N}\Big[\phi_2(2\lambda)-\phi_{2\alpha_-/\eta+1}(\lambda)-
 \phi_{2\alpha_+/\eta+1}(\lambda)+\gamma_+(\lambda)+\gamma_-(\lambda)\Big.\Bigg.\nonumber\\
 &&\Big.\Bigg.+\pi-\sum_{l=1}^{M}(\phi_2(\lambda-\lambda_l)+\phi_2(\lambda+\lambda_l))\Big]\Bigg\}.\label{haiBAEsln}
\end{eqnarray}
In the thermodynamic limit $N\rightarrow\infty$, the distribution of Bethe roots tend to continuous and
\begin{eqnarray}
\frac{dZ(\lambda)}{d\lambda}=\rho(\lambda)+ \rho_h(\lambda),
\end{eqnarray}
where $\rho(\lambda)$ is the density of particles and $\rho_h(\lambda)$ is the density of holes.
From Eq.(\ref{haiBAEsln}), the density of the roots $\rho(\lambda)$ satisfies
\begin{eqnarray}
  \rho(\lambda) &=& \frac{dZ(\lambda)}{d\lambda}
  -\frac{1}{2N}\delta(\lambda)-\frac{1}{2N}\delta(\lambda-\pi)\no \\
   &=& g_1(\lambda)+\frac{1}{2N}\Big[2q(\lambda)-g_{2\alpha_-/
   \eta+1}(\lambda)-g_{2\alpha_+/\eta+1}(\lambda)+h_+(\lambda)+h_-(\lambda)\Big.\nonumber\\
&& \Big.-\delta(\lambda)-\delta(\lambda-\pi)\Big]
-\int_{-\pi}^{\pi}g_2(\lambda-v)\rho(v)dv,
\label{BAEs-sun-1113}
\end{eqnarray}
where
\begin{eqnarray}
  g_m(\lambda) &=& \frac{1}{2\pi}\frac{d\phi_m(\lambda)}{d\lambda}=
  \frac{1}{2\pi}\frac{\sinh(m\eta)}{\cosh(m\eta)-\cos(\lambda)}, \no  \\
  h_{\pm}(\lambda) &=& \frac{1}{2\pi}\frac{d\gamma_{\pm}(\lambda)}{d\lambda}=
  -\frac{1}{2\pi}\frac{\sinh(2\beta_{\pm}+\eta)}{\cos(\lambda)+\cosh(2\beta_{\pm}+\eta)},\no \\
  q(\lambda)&=&g_2(2\lambda).
\end{eqnarray}
In equation~\eqref{BAEs-sun-1113}, the presence of delta-functions is due to the fact that $\lambda_j=0$ and $\lambda_j=\pi$ are the
solutions of~\eqref{BAEsln}, which should be excluded, since they make the wavefunction vanish identically~\cite{SK0}.

\begin{figure}[htbp]
\centering
\begin{minipage}{7cm}
\centering
\includegraphics[scale=0.8]{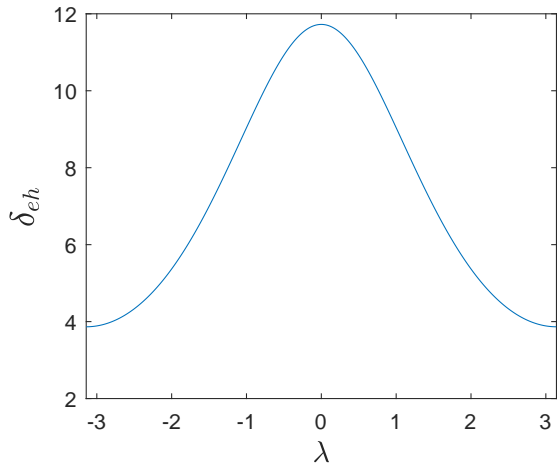}
\end{minipage}
\caption{The energy carried by one bulk hole, where $\eta=2.0$ and $-\pi<\lambda<\pi$. We find $3.8655 < \delta_{eh} < 11.7183$.}
\label{fig:1}
\end{figure}

Now, we consider the elementary excitations of this model.
We first consider the spin excitation, which means that one spin is flipped. The one spin excitation corresponds add two holes in the ground state distribution of $I_{j}$. Denote the positions of holes as
$\lambda_{h}$ and $-\lambda_{h}$.
In the thermodynamic limit $N\rightarrow\infty$, we obtain the density of state $\tilde{\rho}(\lambda)$ in this case is
\begin{eqnarray}
  \tilde{\rho}(\lambda)
   &=& g_1(\lambda)+\frac{1}{2N}\Big[2q(\lambda)-g_{2\alpha_-/
   \eta+1}(\lambda)-g_{2\alpha_+/\eta+1}(\lambda)+h_+(\lambda)+h_-(\lambda)\Big.\nonumber\\
&& \Big.-\delta(\lambda)-\delta(\lambda-\pi)-\delta(\lambda-\lambda_h)-\delta(\lambda+\lambda_h)\Big]
-\int_{-\pi}^{\pi}g_2(\lambda-v)\tilde \rho(v)dv.\label{B2w2AEs-sun-1126}
\end{eqnarray}
From Eqs.(\ref{BAEs-sun-1113}) and (\ref{B2w2AEs-sun-1126}), we obtain the difference between $\tilde{\rho}(\lambda)$ and $\rho(\lambda)$ as $\delta\rho(\lambda)= \tilde{\rho}(\lambda)-\rho(\lambda)$,
which satisfies
\begin{eqnarray}
\delta\rho(\lambda)=\frac{1}{2N}\Big[-\delta(\lambda-\lambda_h)-\delta(\lambda+\lambda_h)
   \Big]-\int_{-\pi}^{\pi}g_2(\lambda-v)\delta\rho(v)dv.
\label{BAEs-sun-1126}
\end{eqnarray}
By using the Fourier transformation
\begin{equation}\label{fly}
  \hat{f}(\omega)=\int_{-\pi}^{\pi}f(\lambda)e^{i\omega\lambda}d\lambda, \qquad
  f(\lambda)=\frac{1}{2\pi}\sum_{\omega=-\infty}^{\infty}\hat{f}(\omega)e^{-i\omega\lambda},
\end{equation}
we obtain the solution of $\delta\rho(\lambda)$ as
\begin{eqnarray}
  \delta\rho(\omega) =-\frac{\cos(\lambda_h\omega)}{N(1+\hat{g}_2(w))},
\end{eqnarray}
where $\hat{g}_m(\omega) = e^{-m\eta|\omega|}$. The energy of a bulk hole at the position $\lambda_{h}$ can be calculated as
\begin{eqnarray}
   \delta_{eh}=2\sinh\eta\sum_{\omega=-\infty}^{\infty}\frac{e^{-iw\lambda_h}}
   {\cosh(\omega\eta)}.
   \label{hole-energy}
 \end{eqnarray}
which is shown in Fig~\ref{fig:1}. The spin of this excitation is $S_z=N\int_{-\pi}^{\pi}\delta \rho(\lambda)d\lambda=-1/2$.

Next, we consider the new solutions of BAEs~\eqref{BAEsre}, that is the boundary strings. The analysis is close to that of~\cite{SK2}.
The fundamental boundary 1-string is the root located at $\lambda_0=2i(\alpha_{\pm}+\frac{\eta}{2})$ for $\alpha_{\pm}>-\frac{\eta}{2}$ and at $\lambda_0=\pi+2i(\beta_{\pm}+\frac{\eta}{2})$
for $\beta_{\pm}>-\frac{\eta}{2}$. One can check that these strings are the solutions of BAEs \eqref{BAEsre}.

\begin{figure}[htbp]
\centering
\begin{minipage}{7cm}
\centering
\includegraphics[scale=0.8]{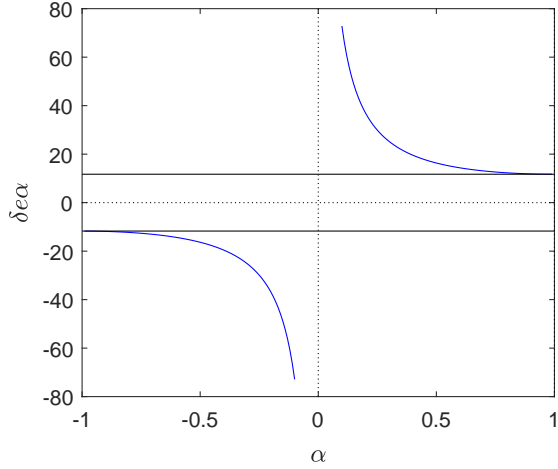}
\end{minipage}
\caption{The energy $\delta_{e\alpha}$ carried by the boundary string located at $\lambda_0=2i(\alpha_\pm+\frac{\eta}{2})$,
where $\eta=2.0$. We find that the absolute value of the energy $\delta_{e\alpha}$ is bigger than the maximum $11.7183$ of the energy $\delta_{eh}$,
$|\delta_{e\alpha}| > \delta^{max}_{eh}=11.7183$ if $-\frac{\eta}2< \alpha_{\pm} < \frac{\eta}2$.}
\label{fig:2}
\end{figure}

Substituting the string solution $\lambda_0=2i(\alpha_{\pm}+\frac{\eta}{2})$
into BAEs \eqref{BAEsre} and taking the thermodynamic limit, we obtain the density of states $\bar {\rho}_\alpha(\lambda)$
\begin{eqnarray}
  \bar{\rho}_\alpha(\lambda)
   &=& g_1(\lambda)+\frac{1}{2N}\Big[2q(\lambda)-g_{2\alpha_-/
   \eta+1}(\lambda)-g_{2\alpha_+/\eta+1}(\lambda)+h_+(\lambda)+h_-(\lambda)\Big.\nonumber\\
&& \Big.-\delta(\lambda)-\delta(\lambda-\pi)
-g_{2}(\lambda-2i(\alpha_{\pm}+\frac{1}{2}\eta))-g_{2}(\lambda+2i(\alpha_{\pm}+\frac{1}{2}\eta))
\Big]\nonumber\\
&&-\int_{-\pi}^{\pi}g_2(\lambda-v)\bar \rho_\alpha(v)dv.   \label{a-ssbae}
\end{eqnarray}
Denote the difference between $\bar{\rho}_{\alpha}(\lambda)$ and $\rho(\lambda)$ as $\delta\rho_{\alpha}(\lambda)= \bar{\rho}_{\alpha}(\lambda)-\rho(\lambda)$.
From Eqs.(\ref{BAEs-sun-1113}) and (\ref{a-ssbae}), we find $\delta\rho_{\alpha}(\lambda)$ should satisfy
\begin{eqnarray}
\delta\rho_\alpha(\lambda)&=& \frac{1}{2N}\Big[-g_{2}(\lambda-2i(\alpha_{\pm}+\frac{1}{2}\eta))-g_{2}(\lambda+2i(\alpha_{\pm}+\frac{1}{2}\eta))
   \Big]\nonumber\\
   &&-\int_{-\pi}^{\pi}g_2(\lambda-v)\delta\rho_\alpha(v)dv.
   \label{a-11sbae}
\end{eqnarray}
The solution of Eq. (\ref{a-11sbae}) is
\begin{eqnarray*}
  \delta\rho_\alpha(\omega) =
  \begin{cases}
   \displaystyle{-\frac{1}{2N(1+\hat{g}_2(\omega))}\left[\hat{g}_{2-2(\alpha_{\pm}+1/2\eta)/\eta}(\omega)
   +\hat{g}_{2+2(\alpha_{\pm}+1/2\eta)/\eta}(\omega)\right], \quad
   -\frac{\eta}{2}<\alpha_\pm<\frac{\eta}{2}},\\[8pt]
   \displaystyle{-\frac{1}{2N(1+\hat{g}_2(\omega))}
   \left[2e^{-2\eta|\omega|}\cosh(2\omega\alpha+\omega\eta)
   -2\cosh(2\omega\alpha-\omega\eta)\right], \quad
   \alpha_\pm>\frac{\eta}{2}}.
   \end{cases}
\end{eqnarray*}
Thus, the energy carried by the boundary string is
\begin{eqnarray}
\label{energy-string-1}
 \delta_{e\alpha} =
  \begin{cases}
   \displaystyle{\frac{4\sinh^2\eta}{\cosh(2\alpha_{\pm}+\eta)-\cosh\eta} } \\[16pt]
   \displaystyle{+2\sinh\eta\sum_{\omega=-\infty}^{\infty}
   \frac{e^{-2\eta|\omega|}\cosh(2\alpha_{\pm}\omega+\eta\omega)
   }{\cosh(\eta\omega)}}, \qquad {-\frac{\eta}{2}<\alpha_\pm<\frac{\eta}{2}},\\[16pt]
   \displaystyle{0, \quad \qquad\alpha_\pm>\frac{\eta}{2}},
   \end{cases}
\end{eqnarray}
which is shown in Fig.~\ref{fig:2}. The corresponding spinor carries the spin $S_z=-1/2$.

\begin{figure}[htbp]
\centering
\begin{minipage}{7cm}
\centering
\includegraphics[scale=0.8]{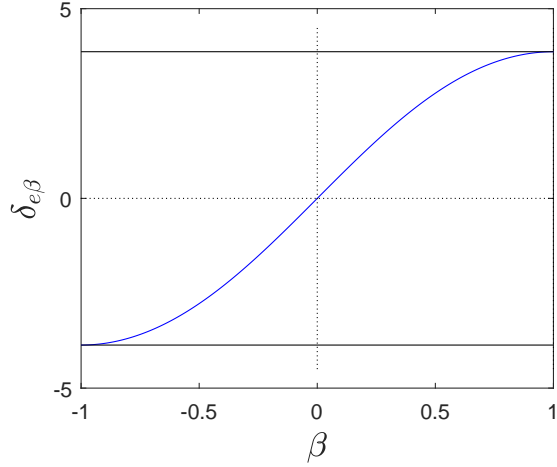}
\end{minipage}
\caption{The energy $\delta_{e\beta}$ carried by the boundary string located at $\lambda_0=\pi+2i(\beta_\pm+\frac{\eta}{2})$,
where $\eta=2.0$. We find that the absolute value of the energy $\delta_{e\beta}$ is smaller than the minimum of the energy $\delta_{eh}$,
$|\delta_{e\beta}| < \delta^{min}_{eh}=3.8655$ if $-\frac{\eta}2< \beta_{\pm} < \frac{\eta}2$.}
\label{fig:3}
\end{figure}

Substituting the string solution $\lambda_0=\pi+2i(\beta_{\pm}+\frac{\eta}{2})$
into BAEs \eqref{BAEsre} and taking the thermodynamic limit, we obtain the density of states $\bar {\rho}_\beta(\lambda)$
\begin{eqnarray}
  \bar{\rho}_\beta(\lambda)
   &=& g_1(\lambda)+\frac{1}{2N}\Big[2q(\lambda)-g_{2\alpha_-/
   \eta+1}(\lambda)-g_{2\alpha_+/\eta+1}(\lambda)+h_+(\lambda)+h_-(\lambda)-\delta(\lambda)\Big.\nonumber\\
&& \Big.-\delta(\lambda-\pi)
-g_{2}(\lambda-\pi-2i(\beta_{\pm}+\frac{1}{2}\eta))-g_{2}(\lambda+\pi+2i(\beta_{\pm}+\frac{1}{2}\eta))
\Big]\nonumber\\
&&-\int_{-\pi}^{\pi}g_2(\lambda-v)\bar \rho_{\beta}(v)dv.
   \label{a-bawwqwe}
\end{eqnarray}
Denote the difference between $\bar{\rho}_{\beta}(\lambda)$ and $\rho(\lambda)$ as $\delta\rho_{\beta}(\lambda)= \bar{\rho}_{\beta}(\lambda)-\rho(\lambda)$.
From Eqs.(\ref{BAEs-sun-1113}) and (\ref{a-bawwqwe}), we find $\delta\rho_{\beta}(\lambda)$ should satisfy
\begin{eqnarray}
\delta\rho_\beta(\lambda)
&=& \frac{1}{2N}\Big[-g_{2}(\lambda-\pi-2i(\beta_{\pm}+\frac{1}{2}\eta))-g_{2}(\lambda+\pi+2i(\beta_{\pm}+\frac{1}{2}\eta))
   \Big]\nonumber\\
   &&-\int_{-\pi}^{\pi}g_2(\lambda-v)\delta\rho_{\beta}(v)dv.
   \label{a-bae}
\end{eqnarray}
Thus, the energy carried by the boundary string is
\begin{eqnarray}
\label{energy-string-2}
 \delta_{e\beta} =
  \begin{cases}
   \displaystyle{-\frac{4\sinh^2\eta}{\cosh(2\beta_{\pm}+\eta)+\cosh\eta}}\\[12pt]
   \displaystyle{+2\sinh\eta\sum_{\omega=-\infty}^{\infty}
   \frac{(-1)^{\omega}e^{-2\eta|\omega|}\cosh(2\beta_{\pm}\omega+\eta\omega)}{\cosh(\eta\omega)},\quad -\frac{\eta}{2}<\beta_{\pm}<\frac{\eta}{2}},\\[12pt]
   0, \displaystyle{\quad \beta_\pm>\frac{\eta}{2}},
   \end{cases}
\end{eqnarray}
which is shown in Fig.~\ref{fig:3}. The corresponding spinor carries the spin $S_z=-1/2$.

Combining the results \eqref{hole-energy},~\eqref{energy-string-1} and~\eqref{energy-string-2}, we find that the excitation energy caused
by the boundary parameter $\alpha_\pm$~\eqref{energy-string-1} is bigger than the maximum of energy of one bulk hole $\delta_{eh}$ \eqref{hole-energy} if
$-\frac{\eta}{2}<\alpha_\pm<\frac{\eta}{2}$, while the
excitation energy caused by the boundary parameter $\beta_\pm$~\eqref{energy-string-2} is smaller than the minimum of energy of one bulk hole~\eqref{hole-energy} if
$-\frac{\eta}{2}<\beta_\pm<\frac{\eta}{2}$. In addition, we conclude that
\begin{eqnarray}
  &&\delta_{e\alpha}<\delta_{e\beta}<0<\delta_{eh};\quad
  -\delta_{e\alpha}>\delta_{eh}>-\delta_{e\beta}, \quad\text{if}
  -\frac{1}{2}\eta<\alpha_{\pm}<0, \quad -\frac{1}{2}\eta<\beta_{\pm}<0,\nonumber\\
  &&\delta_{e\alpha}>\delta_{eh}>\delta_{e\beta}>0, \qquad \text{if}\quad 0<\alpha_{\pm}<\frac{1}{2}\eta,\quad  0<\beta_{\pm}<\frac{1}{2}\eta.
  \label{principle-energy}
\end{eqnarray}

Another conclusion is that the energy of the boundary bound state in the regime of $-\frac{1}{2}\eta<\alpha_{\pm}<0$ is bigger than
the top of the energy band. Therefore it is stable, in spite of its huge energy.

Besides the fundamental boundary 1-string, there exists an infinite set of `long' boundary
strings, consisting of roots $\lambda_0-2ik\eta,\lambda_0-2i(k-1)\eta,\dots,\lambda_0+2ni\eta$ with $n,k\geq0$. We
call such solution an $(n,k)$ boundary string, where (0,0) string is the fundamental boundary
string. By using the same arguments in~\cite{SK3}, we can prove that the $(n,k)$ string
is a solution of BAEs when its `centre of mass' has positive imaginary part and the lowest
root $\lambda_0-2ik\eta$ lies below the real axis. However, a direct calculation shows
that the energy of the $(n,k)$ strings vanishes with $k\geq1$.
For the $(n,0)$ strings with $n\geq1$, they have the same energy as that of the boundary bound
state given by \eqref{energy-string-1} and \eqref{energy-string-2}, so they represent
charged bounary excitations.

\section{Finite size correction}
\label{sec:Finite size correction}
\setcounter{equation}{0}

Now, we consider the contribution of the inhomogeneous term in the $T-Q$ relation (\ref{T-Q}) to the ground state energy of the system.
For this purpose, we define
\begin{equation}\label{Einhom}
 E_{inh}\equiv E_{hom}-E_{true},
\end{equation}
where $E_{true}$ is the ground state energy of the Hamiltonian~\eqref{Hamiltonian} which can be obtained by using the density matrix renormalization group (DMRG)~\cite{DMRG,DMRG1},
while $E_{hom}$ is the minimal energy which can be obtained from~\eqref{Ehom}, where Bethe roots should satisfy the BAEs~\eqref{BAEsln}.
Without losing generality, we choose $\theta_{\pm}=0$, $\alpha_+=\eta$ and $N$ is even.

We analyze the structure of the
Bethe roots at the ground state based on Eq.~\eqref{principle-energy}. For the $(0,0)$ string shown in section~\ref{sec:surface excitations}, the charge
of boundary excitations turned out to be half-integer. We can then conclude that a boundary
excitation can only appear paired with the bulk excitation of half-integer charge or with another
boundary excitation. At the same time, the energy must be smaller than all the real roots.
Let us consider these cases separately.

I. The ground state has no boundary strings.
\begin{table}[htbp]
\centering 
\begin{spacing}{1.15}
\begin{tabular}{|c|c|c|} \hline
${\rm No.}$ & $\text{Regimes of boundary parameters}$ & $\text{Bethe roots}$\\\hline
$1.1$ & $
                                \begin{array}{cccc}
                                  \alpha_+>0, \alpha_->0, \beta_+>\frac{\eta}{2}, \beta_-<-\frac{\eta}{2}  \\
                                \end{array}
$
&  $\frac{N}{2}-1$ \text{real roots} \\
$1.2$ & $                                \begin{array}{cccc}
                                  \alpha_+>0, \alpha_-<-\frac{\eta}{2}, \beta_+>\frac{\eta}{2}, \beta_->\frac{\eta}{2}  \\
                                \end{array}
$
&  $\text{+ one bulk hole}
 $\\\hline
$1.3$ & $                                \begin{array}{cccc}
                                  \alpha_+>0, \alpha_->0, -\frac{\eta}{2} <\beta_+<0, -\beta_+<\beta_-<\frac{\eta}{2}  \\
                                \end{array}
$
&  $$\\
$1.4$ & $                               \begin{array}{cccc}
                                  \alpha_+>0, \alpha_->0, -\frac{\eta}{2}<\beta_+<0, \beta_->\frac{\eta}{2}  \\
                                \end{array}
$
&  $$\\
$1.5$ & $                               \begin{array}{cccc}
                                  \alpha_+>0, \alpha_->0, 0<\beta_+<\frac{\eta}{2}, 0<\beta_-<\frac{\eta}{2}  \\
                                \end{array}
$
&  $\frac{N}{2}-1$ \text{real roots}\\
$1.6$ & $                               \begin{array}{cccc}
                                  \alpha_+>0, \alpha_->0, 0<\beta_+<\frac{\eta}{2}, \beta_->\frac{\eta}{2}  \\
                                \end{array}
$
&  $$\\
$1.7$ & $                               \begin{array}{cccc}
                                  \alpha_+>0, \alpha_->0, \beta_+>\frac{\eta}{2}, \beta_->\frac{\eta}{2}  \\
                                \end{array}
$
&  $$\\\hline
$1.8$ & $                               \begin{array}{cccc}
                                  \alpha_+>0, \alpha_->0, \beta_+<-\frac{\eta}{2}, \beta_-<-\frac{\eta}{2}  \\
                                \end{array}
$
&  $$\\

$1.9$ & $                               \begin{array}{cccc}
                                  \alpha_+>0, \alpha_-<-\frac{\eta}{2}, -\frac{\eta}{2}<\beta_+<0, \beta_-<-\frac{\eta}{2}  \\
                                \end{array}
$
&  $\frac{N}{2}$ \text{real roots}\\
$1.10$ & $                                \begin{array}{cccc}
                                  \alpha_+>0, \alpha_-<-\frac{\eta}{2}, 0<\beta_+<\frac{\eta}{2}, \beta_-<-\frac{\eta}{2}  \\
                                \end{array}
$
&  $$\\
$1.11$ & $                               \begin{array}{cccc}
                                  \alpha_+>0, \alpha_-<-\frac{\eta}{2}, \beta_+>\frac{\eta}{2}, \beta_-<-\frac{\eta}{2}  \\
                                \end{array}
$
&  $$\\\hline

$1.12$ & $                               \begin{array}{cccc}
                                  \alpha_+>0, \alpha_-<-\frac{\eta}{2}, \beta_+<-\frac{\eta}{2}, \beta_-<-\frac{\eta}{2}  \\
                                \end{array}
$
&  $
  \begin{array}{c}
    \frac{N}{2} \quad \text{real roots} \\
    \text{+ one bulk hole} \\
  \end{array}
$\\

\hline
\end{tabular}
\caption{\label{table1} The Bethe roots at the ground state which have no boundary strings.}
\end{spacing}
\end{table}
\begin{figure}[htbp]
\centering
\subfigure[]{
\begin{minipage}{7cm}
\centering
\includegraphics[scale=0.75]{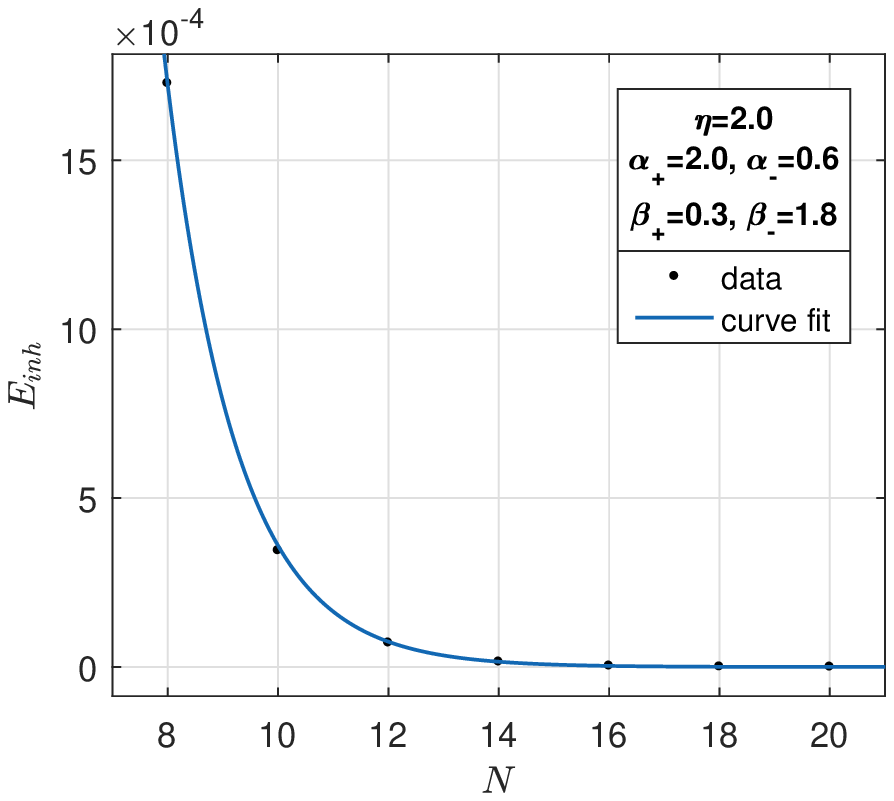}
\end{minipage}}
\subfigure[]{
\begin{minipage}{7cm}
\centering
\includegraphics[scale=0.7]{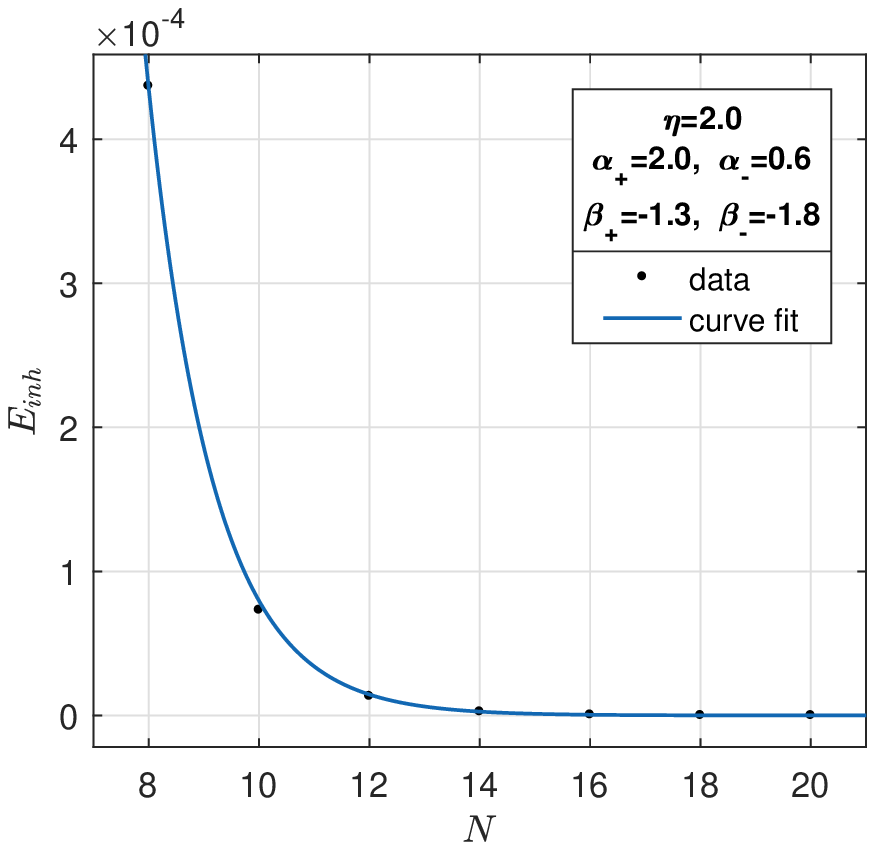}
\end{minipage}}
\caption{The contribution of the inhomogeneous term to the ground state energy $ E_{inh}$ versus the even system-size $N$.
The data can be fitted as $ E_{inh}(N)=p_1e^{q_1N}$.
Here (a) $p_1=0.9139$ and $q_1=-0.7841$;
(b) $p_1=0.3916$ and $q_1=-0.8501$.}
\label{fig:g-1}
\end{figure}

We first consider the case that there is no boundary strings at the ground state. The corresponding regimes of the boundary parameters
are given by Table \ref{table1}. We calculate the energy $E_{inh}$ in these regimes. We find that $E_{inh}$ satisfies the finite-size behavior,
$ E_{inh}(N)=p_1e^{q_1N}$, where $q_1<0$. Which means if $N \rightarrow \infty$, then $E_{inh} \rightarrow 0$. Thus $E_{hom}$ equals to the true
ground state energy in the thermodynamic limit. Without losing generality, Fig.~\ref{fig:g-1} gives the detailed results in the regimes 6 and 8 as the examples.

II. The ground state has one $(0,0)$ string.

\begin{table}[htbp]
\centering 
\begin{spacing}{1.15}
\begin{tabular}{|c|c|c|} \hline
${\rm No.}$ & $\text{Regimes of boundary parameters}$ & $\text{Bethe Roots}$\\\hline
$2.1$ & $                                \begin{array}{cccc}
                                  \alpha_+>0, \alpha_->0, 0<\beta_+<\frac{\eta}{2}, \beta_-<-\frac{\eta}{2}  \\
                                \end{array}
$
&  $
$\\
$2.2$ & $                               \begin{array}{cccc}
                                  \alpha_+>0, \alpha_->0, -\frac{\eta}{2}<\beta_+<0, \beta_-<-\frac{\eta}{2}  \\
                                \end{array}
$
&  $ \frac{N}{2}-1$  \text{real roots} \\
$2.3$ & $                                \begin{array}{cccc}
                                  \alpha_+>0, \alpha_-<-\frac{\eta}{2}, -\frac{\eta}{2}<\beta_+<0, \beta_->\frac{\eta}{2}  \\
                                \end{array}
$
&  $      +  [\pi+2i(\beta_++\frac{\eta}{2})]
$\\
$2.4$ & $                                \begin{array}{cccc}
                                  \alpha_+>0, \alpha_-<-\frac{\eta}{2}, 0<\beta_+<\frac{\eta}{2},  \beta_->\frac{\eta}{2}  \\
                                \end{array}
$&$ $\\
$2.5$ & $                                \begin{array}{cccc}
                                  \alpha_+>0, \alpha_-<-\frac{\eta}{2}, -\frac{\eta}{2}<\beta_+<0, 0<\beta_-<\frac{\eta}{2}  \\
                                \end{array}
$
&  $
$\\\hline
$2.6$ & $                                \begin{array}{cccc}
                                  \alpha_+>0, -\frac{\eta}{2}<\alpha_-<0, 0<\beta_+<\frac{\eta}{2}, \beta_-<-\frac{\eta}{2}  \\
                                \end{array}
$
&  $
$\\
$2.7$ & $                                \begin{array}{cccc}
                                  \alpha_+>0, -\frac{\eta}{2}<\alpha_-<0, -\frac{\eta}{2}<\beta_+<0, \beta_-<-\frac{\eta}{2}  \\
                                \end{array}
$
&  $       \frac{N}{2}-1$ \text{real roots} \\
$2.8$ & $                                \begin{array}{cccc}
                                  \alpha_+>0, -\frac{\eta}{2}<\alpha_-<0, \beta_+>\frac{\eta}{2}, \beta_-<-\frac{\eta}{2}  \\
                                \end{array}
$
&  $ + 2i(\alpha_-+\frac{\eta}{2})
$\\\hline
$2.9$ & $                               \begin{array}{cccc}
                                  \alpha_+>0,  \alpha_-<-\frac{\eta}{2}, -\frac{\eta}{2}<\beta_+<0,  -\frac{\eta}{2}<\beta_-<0  \\
                                \end{array}
$
&  $ \frac{N}{2}-1$  \text{real roots} \\
$2.10$ & $                               \begin{array}{cccc}
                                  \alpha_+>0, \alpha_-<-\frac{\eta}{2}, 0<\beta_+<\frac{\eta}{2}, 0<\beta_-<\frac{\eta}{2}  \\
                                \end{array}
$
&  $      +[ \pi+2i(\text{min}(\beta_\pm)+\frac{\eta}{2})]
$\\\hline

$2.11$ & $                                \begin{array}{cccc}
                                  \alpha_+>0, -\frac{\eta}{2}<\alpha_-<0, \beta_+<-\frac{\eta}{2}, \beta_-<-\frac{\eta}{2}  \\
                                \end{array}
$
&  $
       \begin{array}{c}
         \frac{N}{2}-1 \quad \text{real roots} \\
         \text{+one bulk hole}+2i(\alpha_-+\frac{\eta}{2}) \\
       \end{array}

   $ \\\hline

$2.12$ & $                                \begin{array}{cccc}
                                  \alpha_+>0, -\frac{\eta}{2}<\alpha_-<0, \beta_+>\frac{\eta}{2}, \beta_->\frac{\eta}{2}  \\
                                \end{array}
$
&  $
       \begin{array}{c}
         \frac{N}{2}-2 \quad \text{real roots} \\
         \text{+one bulk hole} +2i(\alpha_-+\frac{\eta}{2}) \\
       \end{array}

   $ \\\hline
\end{tabular}
\caption{\label{table2} The Bethe roots of the ground state which have one (0,0) string.}
\end{spacing}
\end{table}

\begin{figure}[htbp]
\centering
\subfigure[]{
\begin{minipage}{7cm}
\centering
\includegraphics[scale=0.75]{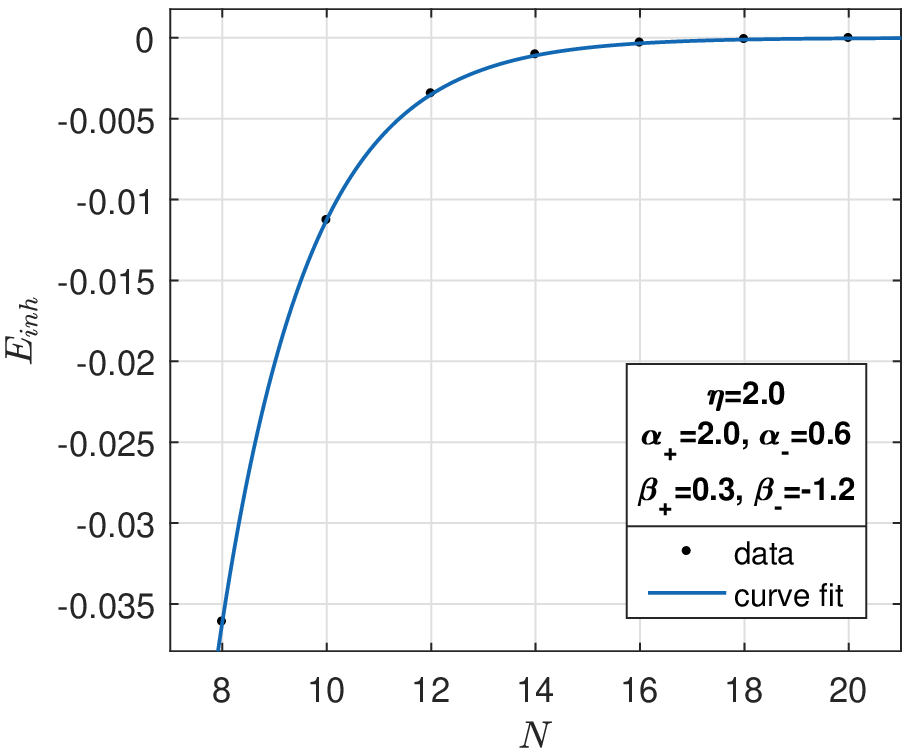}
\end{minipage}}
\subfigure[]{
\begin{minipage}{7cm}
\centering
\includegraphics[scale=0.7]{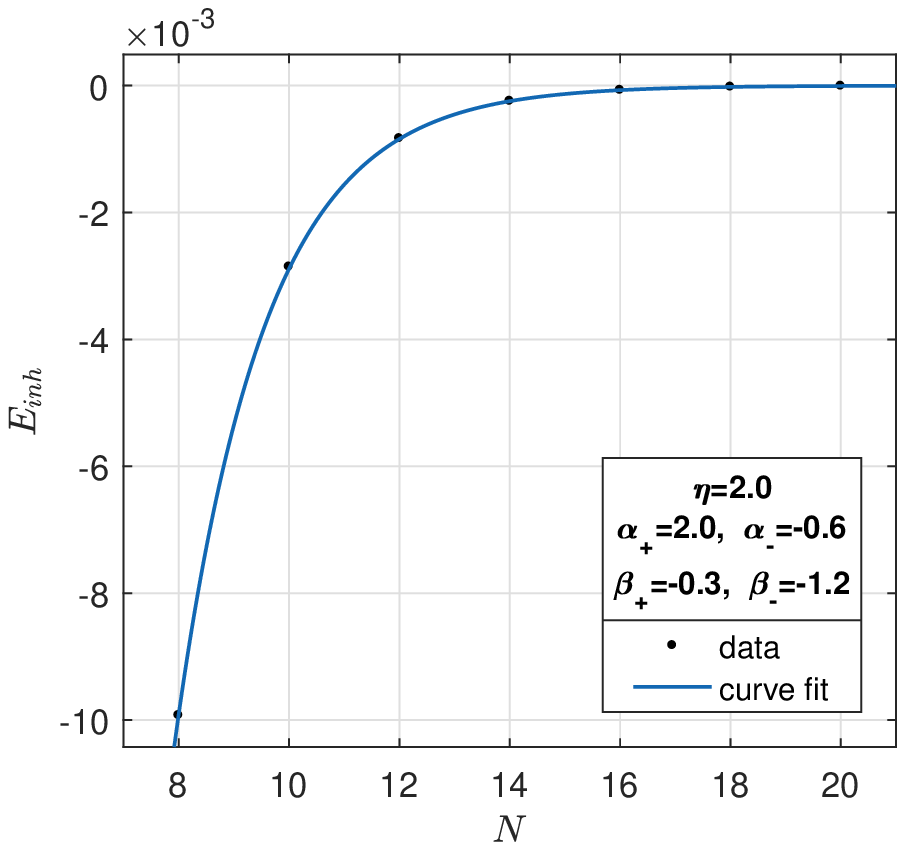}
\end{minipage}}
\caption{The contribution of the inhomogeneous term to the ground state energy $ E_{inh}$ versus the even system-size $N$.
The data can be fitted as $ E_{inh}(N)=p_1e^{q_1N}$.
Here (a) $p_1=-3.8560$ and $q_1=-0.5838$;
(b)$p_1=-1.3750$ and $q_1=-0.6165$.}
\label{fig:g-2}
\end{figure}

Next, we consider the case that there is one boundary strings at the ground state. The corresponding regimes of the boundary parameters
are given by Table \ref{table2} and the finite-size behavior of $E_{inh}$ is shown in Fig. \ref{fig:g-2}. Again, we see that
the inhomogeneous term in (\ref{T-Q}) can be neglected in the thermodynamic limit.

III. The ground state has two $(0,0)$ strings.

\begin{table}[htbp]
\centering 
\begin{spacing}{1.13}
\begin{tabular}{|c|c|c|} \hline
${\rm No.}$ & $\text{Regimes of boundary parameters}$ & $\text{Bethe Roots}$\\\hline
$3.1$ & $                               \begin{array}{cccc}
                                  \alpha_+>0, \alpha_->0, -\frac{\eta}{2}<\beta_+<0, 0<\beta_-<-\beta_+ \\
                                \end{array}
$
&  $ \frac{N}{2}-2$ \text{real roots} \\
$3.2$ & $                                \begin{array}{cccc}
                                  \alpha_+>0, \alpha_->0,  -\frac{\eta}{2}<\beta_+<0, -\frac{\eta}{2}<\beta_-<0 \\
                                \end{array}
$
&  $     \begin{array}{c}
       +[\pi+2i(\beta_-+\frac{\eta}{2})] \\
        +[\pi+2i(\beta_++\frac{\eta}{2})]  \\
      \end{array}
$\\\hline
$3.3$ & $                             \begin{array}{cccc}
                                  \alpha_+>0,  -\frac{\eta}{2}<\alpha_-<0, 0<\beta_+<\frac{\eta}{2}, \beta_->\frac{\eta}{2} \\
                                \end{array}
$
&  $ \frac{N}{2}-2$ \text{real roots} \\
$3.4$ & $                               \begin{array}{cccc}
                                  \alpha_+>0, -\frac{\eta}{2}<\alpha_-<0,  -\frac{\eta}{2}<\beta_+<0, \beta_->\frac{\eta}{2} \\
                                \end{array}
$
&  $     \begin{array}{c}
       +[ \pi+2i(\beta_++\frac{\eta}{2})] \\
      \end{array}
$\\
$3.5$ & $                                \begin{array}{cccc}
                                  \alpha_+>0, -\frac{\eta}{2}<\alpha_-<0, -\frac{\eta}{2}<\beta_+<0, 0<\beta_-<\frac{\eta}{2} \\
                                \end{array}
$
&  $  +[2i(\alpha_-+\frac{\eta}{2})]
$\\\hline
$3.6$ & $                               \begin{array}{cccc}
                                  \alpha_+>0, -\frac{\eta}{2}<\alpha_-<0,  0<\beta_+<\frac{\eta}{2}, 0<\beta_-<\frac{\eta}{2} \\
                                \end{array}
$
&  $  \frac{N}{2}-2$ \text{real roots} \\
$3.7$ & $                               \begin{array}{cccc}
                                  \alpha_+>0, -\frac{\eta}{2}<\alpha_-<0,  -\frac{\eta}{2}<\beta_+<0,  -\frac{\eta}{2}<\beta_-<0 \\
                                \end{array}
$
&  $      \begin{array}{c}
       +[ \pi+2i(\text{min}(\beta_\pm)+\frac{\eta}{2})]  \\
       + 2i(\alpha_-+\frac{\eta}{2})  \\
      \end{array}
$\\
\hline
\end{tabular}
\caption{\label{table3} The Bethe roots of the ground state which have two (0,0) strings.}
\end{spacing}
\end{table}

\begin{figure}[htbp]
\centering
\subfigure[]{
\begin{minipage}{7cm}
\centering
\includegraphics[scale=0.7]{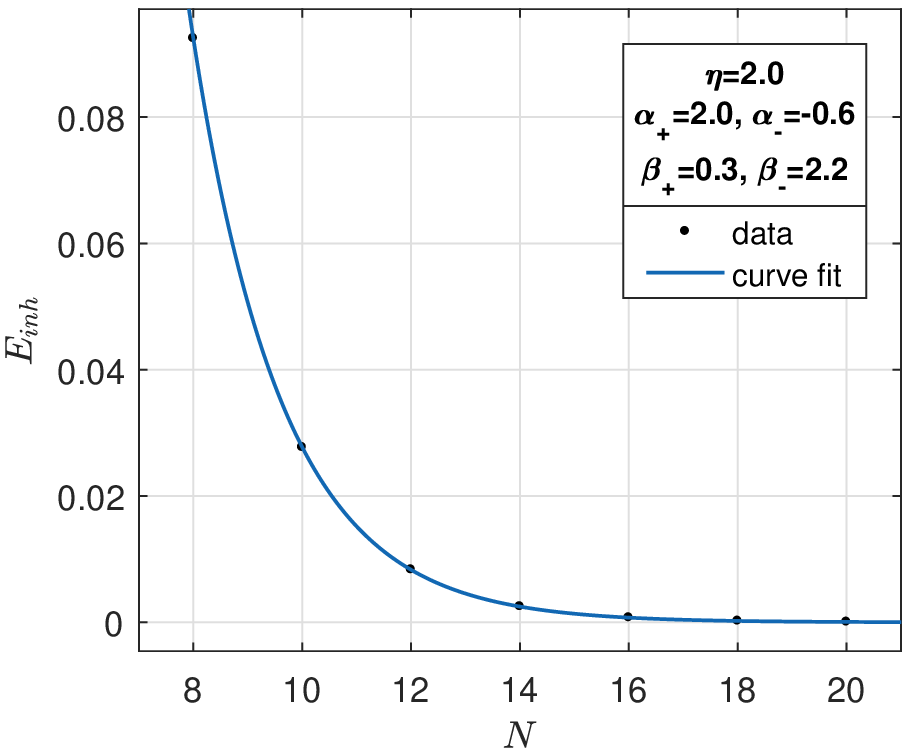}
\end{minipage}}
\subfigure[]{
\begin{minipage}{7cm}
\centering
\includegraphics[scale=0.68]{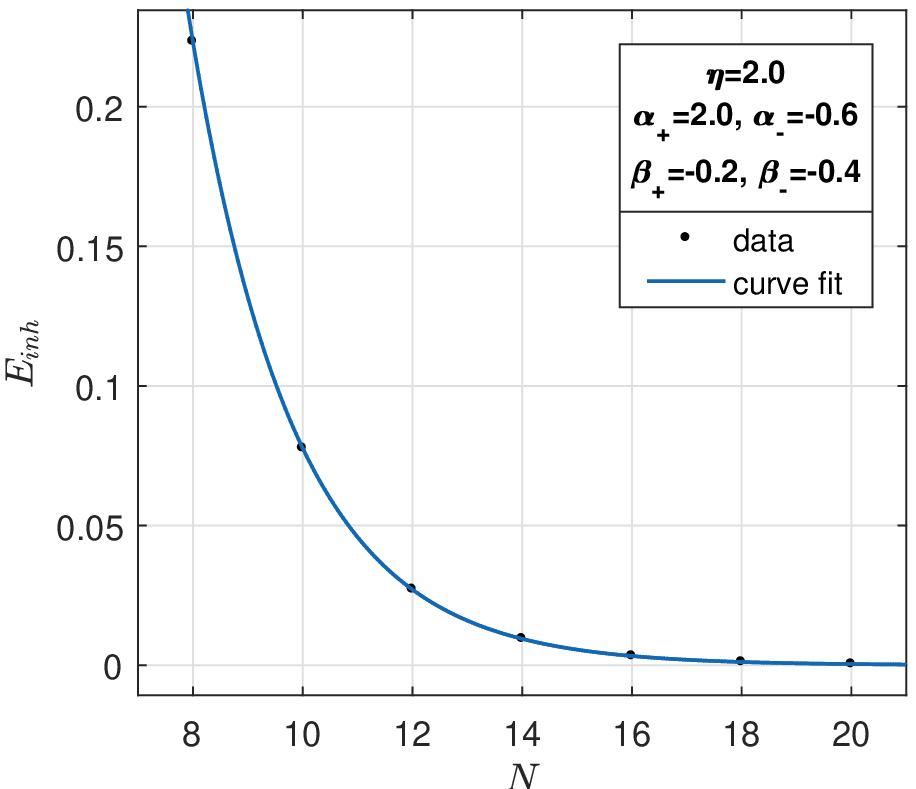}
\end{minipage}}
\caption{The contribution of the inhomogeneous term to the ground state energy $ E_{inh}$ versus the even system-size $N$.
The data can be fitted as $ E_{inh}(N)=p_1e^{q_1N}$.
Here (a) $p_1=11.4200$ and $q_1=-0.6020$; (b) $p_1=15.1900$ and $q_1=-0.5275$.}
\label{fig:g-3}
\end{figure}

Last, we consider the case that there are two boundary strings at the ground state. The corresponding regimes of the boundary parameters
are given by Table \ref{table3} and the finite-size behavior of $E_{inh}$ is shown in Fig. \ref{fig:g-3}. Again, we see that
the inhomogeneous term in the $T-Q$ relation (\ref{T-Q}) can be neglected in the thermodynamic limit.

\section{Surface energy}
\label{sec:surface energy}
\setcounter{equation}{0}

Now we consider the surface energy induced by the boundary magnetic fields.
For the condition that shown in Table (\ref{table1}), in which all the Bethe roots are real at the ground state.
Taking the Fourier
transformation of equation~\eqref{BAEs-sun-1113}, we obtain
\begin{equation}\label{}
  \hat{\rho}(\omega)=\hat{\rho}_0(\omega)+\hat{\rho}_b^0(\omega)
  +\hat{\rho}_{\beta_+}(\omega)
  +\hat{\rho}_{\beta_-}(\omega)+\hat{\rho}_{\alpha_+}(\omega)
  +\hat{\rho}_{\alpha_-}(\omega),
\end{equation}
where
\begin{eqnarray}
\hat{\rho}_b^0(\omega)&=&\frac{2\hat{q}(\omega)-1-(-1)^w}{2N(1+\hat{g}_2(\omega))}, \qquad
   \hat{q}(\omega) = \frac{e^{-\eta|\omega|}}{2}(1+(-1)^{\omega}), \no \\[8pt]
   \hat{\rho}_{\alpha_{\pm}}(\omega) &=& \left\{
\begin{array}{cc}
\displaystyle{   -\frac{\hat{g}_{(2\alpha_{\pm}/\eta+1)}(\omega)}{2N(1+\hat{g}_2(\omega))}}, &
   \quad \displaystyle{ \alpha_{\pm}>-\frac{\eta}{2},} \\[8pt]
\displaystyle{  \frac{\hat{g}_{(-2\alpha_{\pm}/\eta-1)}(\omega)}{2N(1+\hat{g}_2(\omega))}}, &
   \quad \displaystyle{ \alpha_{\pm}<-\frac{\eta}{2},}
\end{array} \right.   \no \\[8pt]
   \hat{\rho}_{\beta_{\pm}}(\omega) &=&\left\{
\begin{array}{cc}
 \displaystyle{    \frac{\hat{h}_{\pm}(\omega)}{2N(1+\hat{g}_2(\omega))}}, & \quad \displaystyle{   \beta_{\pm}< -\frac{\eta}{2}}, \no \\
\displaystyle{     -\frac{\hat{h}_{\pm}(\omega)}{2N(1+\hat{g}_2(\omega))}}, & \quad \displaystyle{   \beta_{\pm}>-\frac{\eta}{2}},
\end{array} \right.   \no \\[8pt]
   \hat{h}_{\pm}(\omega)  &=&  (-1)^{\omega}e^{-2|\beta_{\pm}\omega|-|\eta \omega|}, \quad \hat{\rho}_0(\omega)=\frac{1}{2\cosh(\eta\omega)}.
\end{eqnarray}
The ground energy can be expressed as
\begin{eqnarray}
  E &=& -8\pi N \sinh(\eta)\int_{-\pi}^{\pi}g_1(\lambda)\rho(\lambda)d\lambda+N\cosh(\eta)+E_0
  \nonumber\\
  &=& N e_g+e_b,
  \label{ground-energy}
\end{eqnarray}
where
\begin{eqnarray}
  e_b   &=&   e_b^0+e_{\alpha_+}+e_{\alpha_-}+e_{\beta_+}+e_{\beta_-},  \\[4pt]
  e_g   &=&   -2\sinh(\eta)\sum_{\omega=-\infty}^{\infty}
  \frac{e^{-\eta|\omega|}}{\cosh(\eta \omega)} +\cosh(\eta),\no \\[4pt]
  e_b^0   &=&   -\cosh(\eta)-\sum_{\omega=-\infty}^{\infty}
  \frac{[2\tilde{q}(\omega)-1-(-1)^w]\sinh(\eta)}{\cosh(\eta \omega)},\no \\[8pt]
  e_{\alpha_{\pm}}  &=&  \left\{
\begin{array}{cc}
   -\sinh(\eta)\coth(\alpha_{\pm})+\sinh(\eta)
\displaystyle{    \sum_{\omega=-\infty}^{\infty}
  \frac{\tilde{g}_{2\alpha_{\pm}/\eta+1}(\omega)}{\cosh(\eta \omega)}}, & \quad
  \alpha_{\pm}>-\frac{\eta}{2},\no \\
   -\sinh(\eta)\coth(\alpha_{\pm})-\sinh(\eta)
\displaystyle{    \sum_{\omega=-\infty}^{\infty}
  \frac{\tilde{g}_{(-2\alpha_{\pm}/\eta-1)}(\omega)}{\cosh(\eta \omega)}}, & \quad
  \alpha_{\pm}<-\frac{\eta}{2},
  \end{array} \right.   \no \\[8pt]
  e_{\beta_{\pm}}  &=& \left\{
\begin{array}{cc}
   -\sinh(\eta)\tanh(\beta_{\pm})-\sinh(\eta)
\displaystyle{    \sum_{\omega=-\infty}^{\infty}
  \frac{\tilde{h}_{\pm}(\omega)}{\cosh(\eta \omega)}}, &\quad \beta_{\pm}<-\frac{\eta}{2}, \no \\[4pt]
 -\sinh(\eta)\tanh(\beta_{\pm})+\sinh(\eta)
\displaystyle{    \sum_{\omega=-\infty}^{\infty}
  \frac{\tilde{h}_{\pm}(\omega)}{\cosh(\eta \omega)}},& \quad \beta_{\pm}>-\frac{\eta}{2}.
    \end{array} \right.
\end{eqnarray}
Here $e_g$ equals to the ground state energy density of the periodic chain and $e_b$ is the surface energy induced by the open boundary and the boundary fields.

 It's easy to show that for the other conditions, which includes the one boundary $(0,0)$ string and two boundary
 $(0,0)$ strings. The ground state energy can be expressed by two parts. One of them comes from
 the real roots~\eqref{ground-energy} and the other comes from the bulk holes~(\ref{hole-energy}) or the boundary bound strings (\ref{energy-string-1}) - (\ref{energy-string-2})\footnote{The surface energy of this model with special boundary parameters $(\alpha_{\pm}=\alpha,\beta_-=\beta,\beta_+=-\beta)$ for a real $\eta$ has been studied by the quantum transfer matrix method \cite{klumper} in \cite{poli_last1,poli_last}.}.

For simplicity, here we only give two examples.

I. For the interval that the Bethe roots of the ground state are $\frac{N}{2}-1$ real roots plus one $(0,0)$ string, in the regime of
$\alpha_+>0$, $\alpha_->0$, $0<\beta_+<\frac{\eta}{2}$ and $\beta_-<-\frac{\eta}{2}$,
the ground state energy can be expressed by
 \bea
E=N e_g+e_b+\delta_{e\beta_+},
\eea
where $e_b+\delta_{e\beta_+}$ is the surface energy induced by the open boundary and the boundary fields.

II. For the interval that the Bethe roots of the ground state are $\frac{N}{2}-2$ real roots plus two $(0,0)$ strings,
in the regime of $\alpha_+>0$, $\alpha_->0$, $-\frac{\eta}{2}<\beta_+<0$ and $-\frac{\eta}{2}<\beta_-<0$,
the ground state energy is
\bea
E=N e_g+e_b+\delta_{e\beta_+}+\delta_{e\beta_-},
\eea
where $e_b+\delta_{e\beta_+}+\delta_{e\beta_-}$ is the surface energy induced by the open boundary and the boundary fields.

\section{Elementary excitation}
\label{sec:Elementary excitation-anti}
\setcounter{equation}{0}

Now, we consider the elementary excitation. First, we show that the inhomogeneous term in
the $T-Q$ relation~\eqref{T-Q} can also be neglected in the thermodynamic limit for the excited states.
For this purpose, we define
\begin{equation}\label{Einhwqom}
\Delta E=\Delta E^{ED}-\delta_e,
\end{equation}
where $\Delta E^{ED}$ is the minimal change of energy between the ground state and the excitations of the Hamiltonian~\eqref{Hamiltonian}
which can be obtained by using the DMRG. Let $\delta_{e}$ be the minimal change of energy  from the ground state obtaining from~\eqref{Ehom} and~\eqref{BAEsln}.
From the equation~\eqref{principle-energy}, we know that the
energy change $\delta_{e}$  are connected with the choice of boundary parameters. Let us consider them one by one.

I. The ground state has no boundary strings.

\begin{figure}[htbp]
\centering
\subfigure[]{
\begin{minipage}{7cm}
\centering
\includegraphics[scale=0.75]{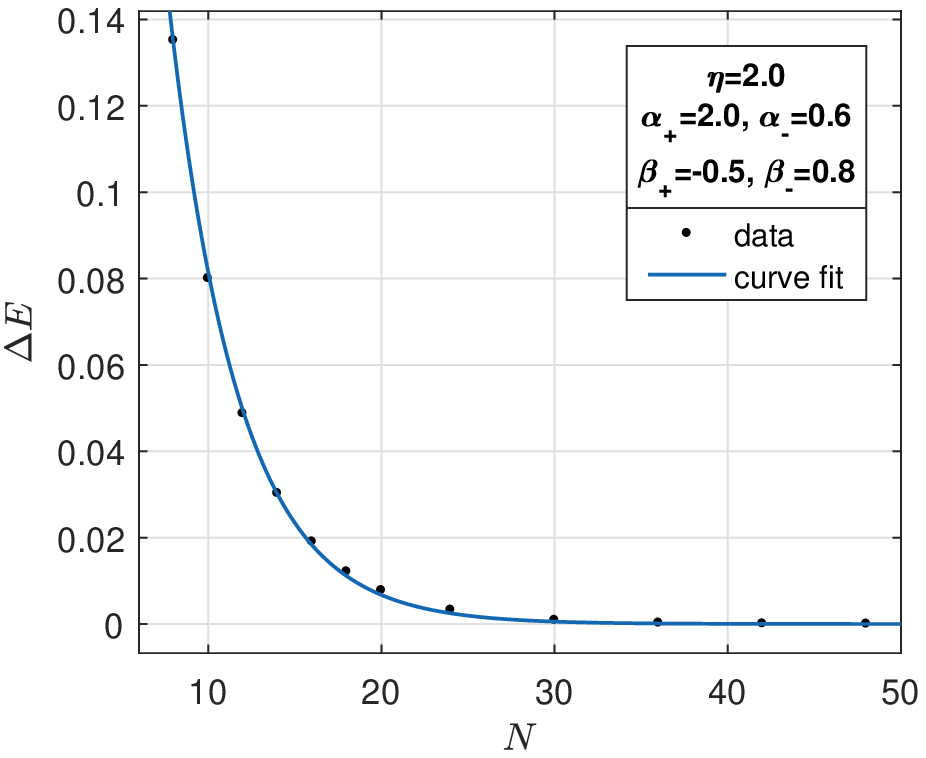}
\end{minipage}}
\subfigure[]{
\begin{minipage}{7cm}
\centering
\includegraphics[scale=0.7]{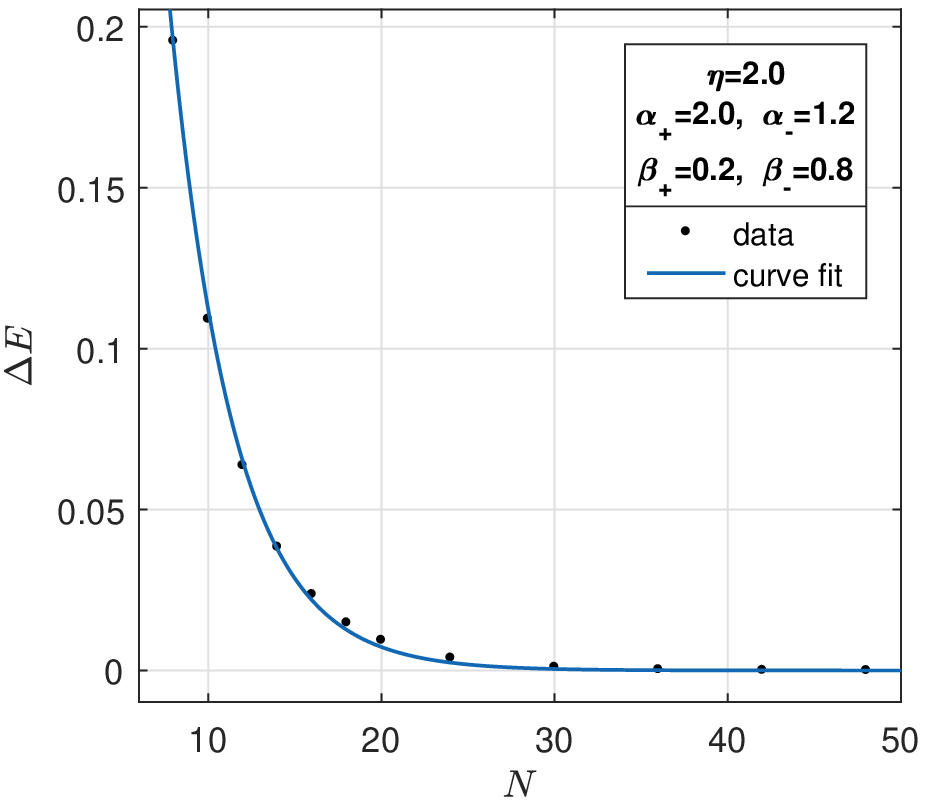}
\end{minipage}}
\caption{$\Delta E=\Delta E^{ED}-\delta e$, where $\Delta E^{ED}$ the minimal change of energy between the ground state and the excitations calculated by using DMRG.
The figure can be fitted as $E(N)=p_2 e^{q_2 N}$.
Here (a) $p_2=0.9875$ and $q_2=-0.2494$;
(b) $p_2=1.7260$ and $q_2=-0.2733$.}
\label{fig:gap-1}
\end{figure}

The finite-size behaviors of $\Delta E$ in the regimes of 1.3 and 1.5 are shown in Fig.~\ref{fig:gap-1}. The fitted curves gives
$\Delta E=p_2 e^{q_2N}$, where $q_2<0$. Thus the $\Delta E$ tends to zero exponentially when the size of the system tends to infinity, and $\delta_{e}$ gives the the minimal change of energy from the ground state in the thermodynamic limit.
The energy change in the whole regimes are given by Table \ref{table4}.
\begin{table}[htbp]
\centering 
\begin{spacing}{1.13}
\begin{tabular}{|c|c|} \hline
{\rm Vaule of} $\delta_e$ & $\text{Regimes of boundary parameters in Table \ref{table1}}$\\\hline
$\delta_{e\beta_+}+\delta_{eh}$ & $1.4, 1.6, 1.9, 1.10
$
\\\hline
$\delta_{eh_1}+\delta_{eh_2}$ & $1.1, 1.2, 1.7, 1.8, 1.11, 1.12$
\\\hline
$\delta_{e\beta_+}+\delta_{e\beta_-}$ & $1.3, 1.5 $
\\
\hline
\end{tabular}
\caption{\label{table4} The Bethe roots at the ground state which have no boundary strings.}
\end{spacing}
\end{table}

II. The ground state contains one boundary $(0,0)$ string.

\begin{figure}[htbp]
\centering
\subfigure[]{
\begin{minipage}{7cm}
\centering
\includegraphics[scale=0.73]{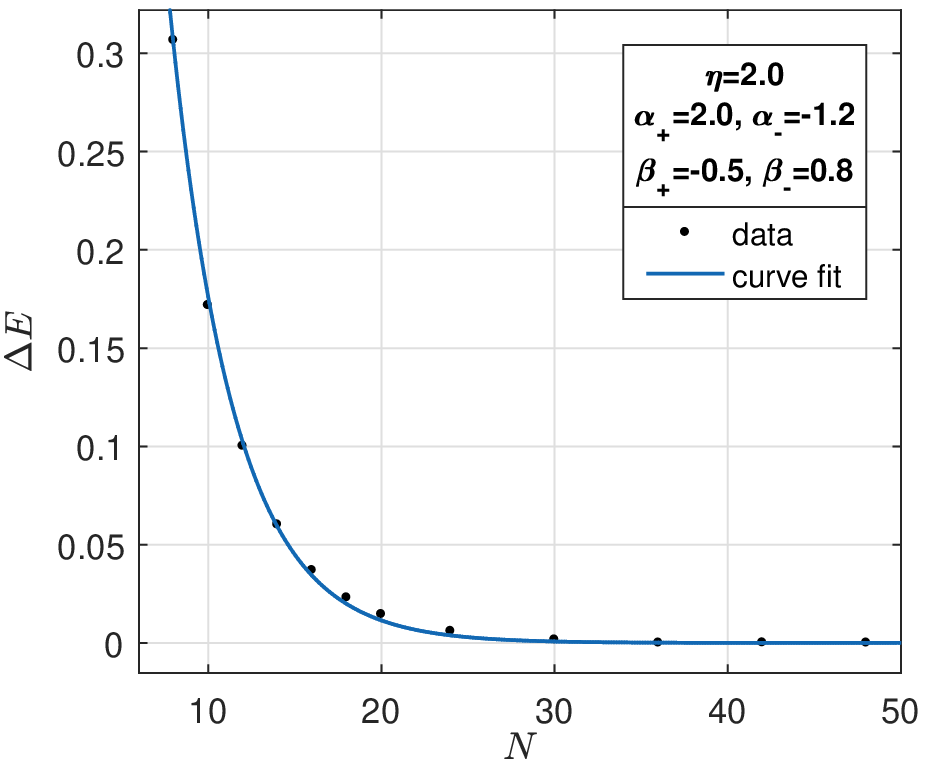}
\end{minipage}}
\subfigure[]{
\begin{minipage}{7cm}
\centering
\includegraphics[scale=0.73]{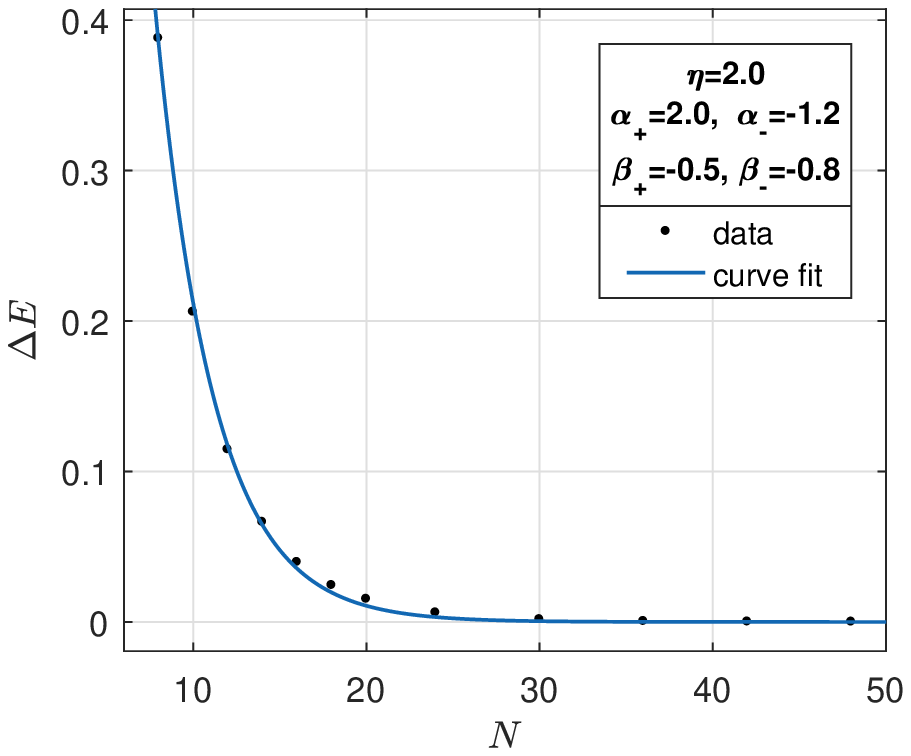}
\end{minipage}}
\caption{$\Delta E=\Delta E^{ED}-\delta e$, where $\Delta E^{ED}$ is the minimal change of energy between the ground state and the excitations calculated by using DMRG. The figure can be fitted as $E(N)=p_2 e^{q_2 N}$.
Here (a) $p_2=2.7060$ and $q_2=-0.2732$;
 (b) $p_2=4.1750$ and $q_2=-0.2980$.}
\label{fig:gap-2}
\end{figure}

The finite-size behaviors of $\Delta E$ in the regimes of 2.5 and 2.9 are shown in Fig.~\ref{fig:gap-2}.
Again, we see that the $\Delta E$ tends to zero and $\delta_{e}$ gives the the minimal change of energy from the ground state in the thermodynamic limit.
The energy change are given by Table \ref{table5}.
\begin{table}[htbp]
\centering 
\begin{spacing}{1.13}
\begin{tabular}{|c|c|} \hline
{\rm Vaule of} $\delta_e$ & $\text{Regimes of boundary parameters in Table \ref{table2}}$\\\hline
$\delta_{e\beta_+}+\delta_{eh}$ & $ 2.6, 2.7
$
\\\hline
$\delta_{eh_1}+\delta_{eh_2}$ & $  2.8, 2.11, 2.12
$
\\\hline
$\delta_{e\beta_-}-\delta_{e\beta_+}$ & $2.5
$
\\\hline
$-\delta_{e\beta_+}+\delta_{eh}$ & $ 2.1, 2.2, 2.3, 2.4
$
\\\hline
$|\delta_{e\beta_+}-\delta_{e\beta_-}|$ & $ 2.9, 2.10
$
\\
\hline
\end{tabular}
\caption{\label{table5}The Bethe roots at the ground state which have one $(0,0)$ string.}
\end{spacing}
\end{table}

III. The ground state contains two boundary $(0,0)$ strings.

\begin{figure}[htbp]
\centering
\subfigure[]{
\begin{minipage}{7cm}
\centering
\includegraphics[scale=0.7]{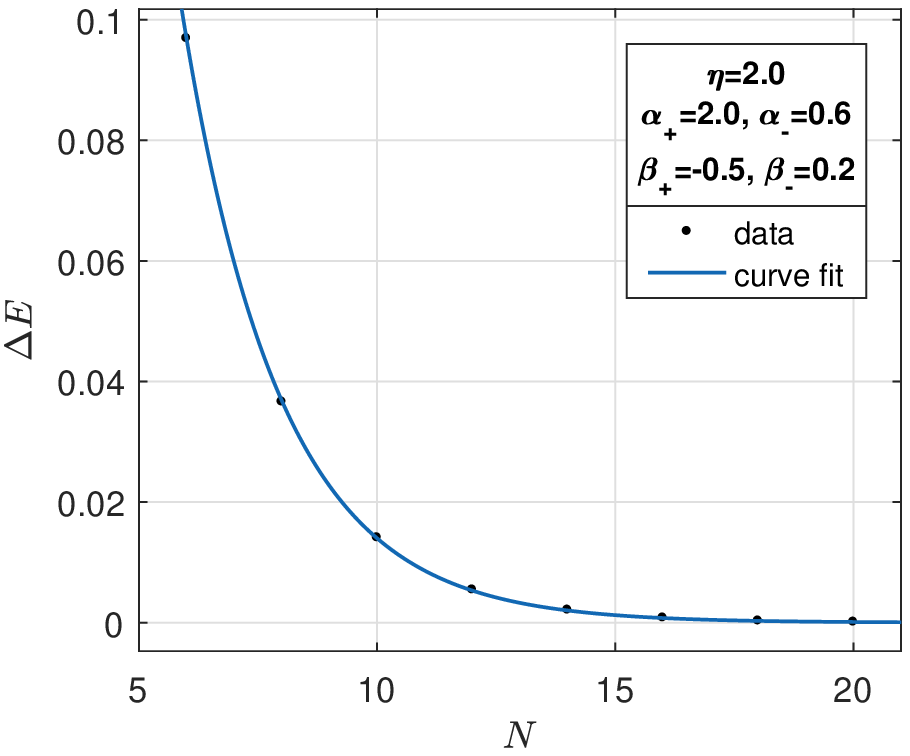}
\end{minipage}}
\subfigure[]{
\begin{minipage}{7cm}
\centering
\includegraphics[scale=0.7]{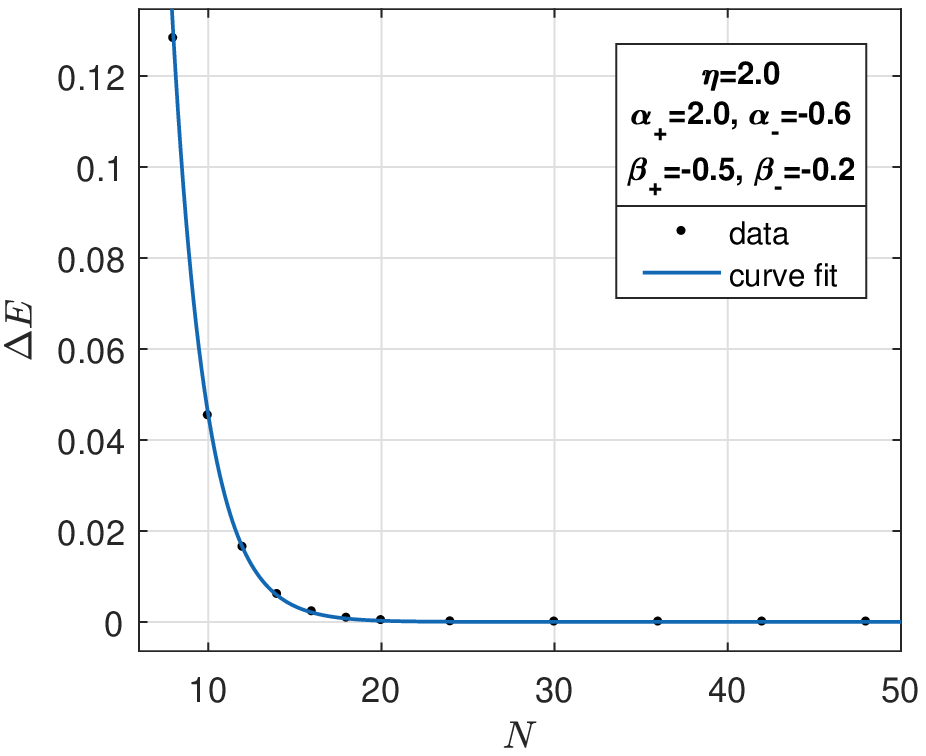}
\end{minipage}}
\caption{$\Delta E=\Delta E^{ED}-\delta e, $ where $\Delta E^{ED}$ is the minimal change of energy between the ground state and the excitations calculated by using DMRG. The figure can be fitted as $E(N)=p_2 e^{q_2 N}$.
Here (a) $p_2=1.7640$ and $q_2=-0.4836$;
(b) $p_2=7.9740$ and $q_2=-0.5163$.}
\label{fig:gap-3}
\end{figure}

The finite-size behaviors of $\Delta E$ in the regimes 3.1 and 3.7 are shown in Fig.~\ref{fig:gap-3} and the energy change are given by Table \ref{table6}.
\begin{table}[htbp]
\centering 
\begin{spacing}{1.13}
\begin{tabular}{|c|c|} \hline
{\rm Vaule of} $\delta_e$ & $\text{Regimes of boundary parameters in Table \ref{table3}}$\\\hline
$\delta_{eh}-\delta_{e\beta_+} $ & $ 3.3, 3.4
$
\\\hline
$\delta_{e\beta_-}-\delta_{e\beta_+}$ & $ 3.5
$
\\\hline
$|\delta_{e\beta_-}-\delta_{e\beta_+}|$ & $ 3.6, 3.7
$
\\\hline
$-\delta_{e\beta_+}-\delta_{e\beta_-} $ & $ 3.1, 3.2
$
\\\hline
\end{tabular}
\caption{\label{table6} The Bethe roots at the ground state which have two$(0,0)$ strings.}
\end{spacing}
\end{table}

\section{Conclusions}
\label{sec:concluding remarks}

In this paper, we study the thermodynamic properties of one-dimensional XXZ spin chain with unparallel boundary magnetic fields at the gaped region ($\eta$ being a real number). Firstly, we analyse the change of energy comes from the bulk hole and the boundary strings of the reduced $T-Q$ relation.
Then we give the distribution of the Bethe roots in the reduced BAEs for different boundary parameters.
Secondly, it is shown that the contribution of the inhomogeneous term in the $T-Q$ relation for the ground state or for the elementary excitation states both can be neglected
when the size of the system $N$ tends to infinity. This allows us to obtain the surface energy and the elementary excitation of the model.

\section*{Acknowledgments}

We would like to thank Prof. Y. Wang for his valuable discussions and continuous encouragements.
The financial supports from the National Program
for Basic Research of MOST (Grant Nos. 2016YFA0300600 and
2016YFA0302104), the National Natural Science Foundation of China
(Grant Nos. 11434013, 11425522, 11547045, 11774397, 11775178 and 11775177), the Major Basic Research Program of Natural Science of Shaanxi Province
(Grant Nos. 2017KCT-12, 2017ZDJC-32), Australian Research Council (Grant No. DP 190101529) and the Strategic Priority Research Program of the Chinese
Academy of Sciences, and the Double First-Class University Construction Project of Northwest University are gratefully acknowledged.
P. Sun is also partially supported by the NWU graduate student innovation funds No. YZZ15088, and would like to thank Dr. F. K. Wen and B. Pozsgay for their
hlepful discussions.

\end{document}